\documentclass[a4paper,10pt,notitlepage]{revtex4-1}

\usepackage{fullpage}
\usepackage{graphicx}
\usepackage{subfigure}
\usepackage[titletoc]{appendix}
\usepackage{epsfig}
\usepackage{dcolumn}
\usepackage{bm}
\usepackage{amsmath,amssymb,color}
\usepackage{setspace}
\begin{document}

\newcommand{\nwc}{\newcommand}
\nwc{\vs}{\vspace}
\nwc{\hs}{\hspace}
\nwc{\la}{\langle}
\nwc{\ra}{\rangle}
\nwc{\lw}{\linewidth}
\nwc{\nn}{\nonumber}
\nwc{\tb}{\textbf}
\nwc{\td}{\tilde}
\nwc{\Tr}{\tb{Tr}}
\nwc{\dg}{\dagger}

\nwc{\pd}[2]{\frac{\partial #1}{\partial #2}}
\nwc{\zprl}[3]{Phys. Rev. Lett. ~{\bf #1},~#2~(#3)}
\nwc{\zpre}[3]{Phys. Rev. E ~{\bf #1},~#2~(#3)}
\nwc{\zpra}[3]{Phys. Rev. A ~{\bf #1},~#2~(#3)}
\nwc{\zjsm}[3]{J. Stat. Mech. ~{\bf #1},~#2~(#3)}
\nwc{\zepjb}[3]{Eur. Phys. J. B ~{\bf #1},~#2~(#3)}
\nwc{\zepjd}[3]{Eur. Phys. J. D ~{\bf #1},~#2~(#3)}
\nwc{\zrmp}[3]{Rev. Mod. Phys. ~{\bf #1},~#2~(#3)}
\nwc{\zepl}[3]{Europhys. Lett. ~{\bf #1},~#2~(#3)}
\nwc{\zjsp}[3]{J. Stat. Phys. ~{\bf #1},~#2~(#3)}
\nwc{\zptps}[3]{Prog. Theor. Phys. Suppl. ~{\bf #1},~#2~(#3)}
\nwc{\zpt}[3]{Physics Today ~{\bf #1},~#2~(#3)}
\nwc{\zap}[3]{Adv. Phys. ~{\bf #1},~#2~(#3)}
\nwc{\zjpcm}[3]{J. Phys. Condens. Matter ~{\bf #1},~#2~(#3)}
\nwc{\zjpa}[3]{J. Phys. A: Math theor  ~{\bf #1},~#2~(#3)}

\title{Maxwell's Demon, Szilard Engine and Landauer Principle}
\author{P. S. Pal$^1$ and A. M. Jayannavar$^{2,3}$}
\email{ pspal@umbc.edu}
\email{ jayan@iopb.res.in}
\affiliation{ $^1$University of Maryland, Baltimore County, 1000 Hilltop Circle
Baltimore, MD 21250 \\ $^2$Institute of Physics, Sachivalaya Marg, Bhubaneswar 751005, India.
\\$^3$Homi Bhabha National Institute, Training School Complex, Anushakti Nagar, Mumbai 400085, India. }
\begin{abstract}
The second law of thermodynamics is probabilistic in nature. Its formulation requires that the state of a system be described by a probability distribution. A natural question, thereby, arises as to whether a prior knowledge about the state of the system affects the second law. This question has now been nurtured over a century and its inception was done by C. Maxwell through his famous thought experiment wherein comes the idea of Maxwell's demon. The next important step in this direction was provided by L. Szilard who demonstrated a theoretical model for an information engine incorporating Maxwell's demon. The final step that lead to the inter-linkage between information theory and thermodynamics was through Landauer's principle of information erasure that established the fact that \textbf{\emph{information is physical}}. Here we will present an overview of these three major works that laid the foundations of information thermodynamics.
\end{abstract}
\maketitle{}
\section*{Introduction} 
Almost 150 years ago, Maxwell put forward a thought experiment thereby threatening the validation of second law of thermodynamics. He introduced a demon like creature which controls the dynamics of an isolated system and while doing so the entropy of the  system decreases\cite{max,vedral_09}. This violates the second law. A classical analysis of Maxwell's demon was conducted by Szilard \cite{szilard} where he studied an idealized heat engine with one particle gas and directly associated  the information acquired by measurement with the physical entropy and saved the second law. Later Landauer put forward his famous principle\cite{landauer_61,landauer_91} which states that it is information erasure rather than information acquired that is associated with an energetic and hence entropic cost. Now it is a widely known fact that the total system consisting of SZE, heat reservoir and demon's memory do not violate the second law. 

Introduction of Maxwell's demon has stirred the physics community working in information themodynamics  and has lead to some insightful physics \cite{Esposito_11,mandal_12,sagawa_12,mandal_13,parrondo_15,rana_16,rana_1611,kim_11,Lu_12,kim_16}. 
 There are intensive interests in quantum thermodynamics of various engines\cite{yan_12,yan_121,wang_12,huang_13,li_13,yuan_14,zhuang_14,huang_14,wang_15}. These microscopic engines are the basis of a larger subject of study, quantum biology, for instance, the avian compass\cite{pearson_16}, and quantum effect in living cells\cite{li_16}.
 
A generalisation of the Szilard engine with multi-particle system as working substance is discussed here. The work extraction in  classical multi-particle Szilard engine will depend upon initial biasing. Szilard engine is generally studied with the partition being initially put in the middle of the box, i.e., the partition divides the phase space of the working substance into two equal halves. Here this division of phase space is made arbitrary by inserting the partition at an arbitrary distance from the boundary. This action changes the probability of particles to stay in certain volume. In this  sense we are calling it a biasing. This affects the work extraction. Another important question would be how this biasing can lead to an optimal work extraction in multi particle Szilard engine. We find that work extraction can be made larger for $N$-particle engine when the partition is inserted near the boundary. 
Another important factor that affects the work extraction is the errors that happen during measurement procedure. Erroneous measurements in general decrease performance as expected.
 

 
\section{Maxwell's Demon}
Suppose, we have two objects at different temperatures $T_1$ and $T_2$$(T_1< T_2)$. If they put in contact with each other, heat will flow from the hot body to the cold one, thereby establishing an equilibrium situation with the final temperature being $T_{\text{fin}}$$(T_1< T_{\text{fin}}< T_2)$. The fact that the temperature of the cold bath cannot decrease and the temperature of the hot bath cannot increase - is prohibited by the second law of thermodynamics.
\begin{figure}[h]
 \vspace{0.5cm}
\centering
\includegraphics[width=7 cm]{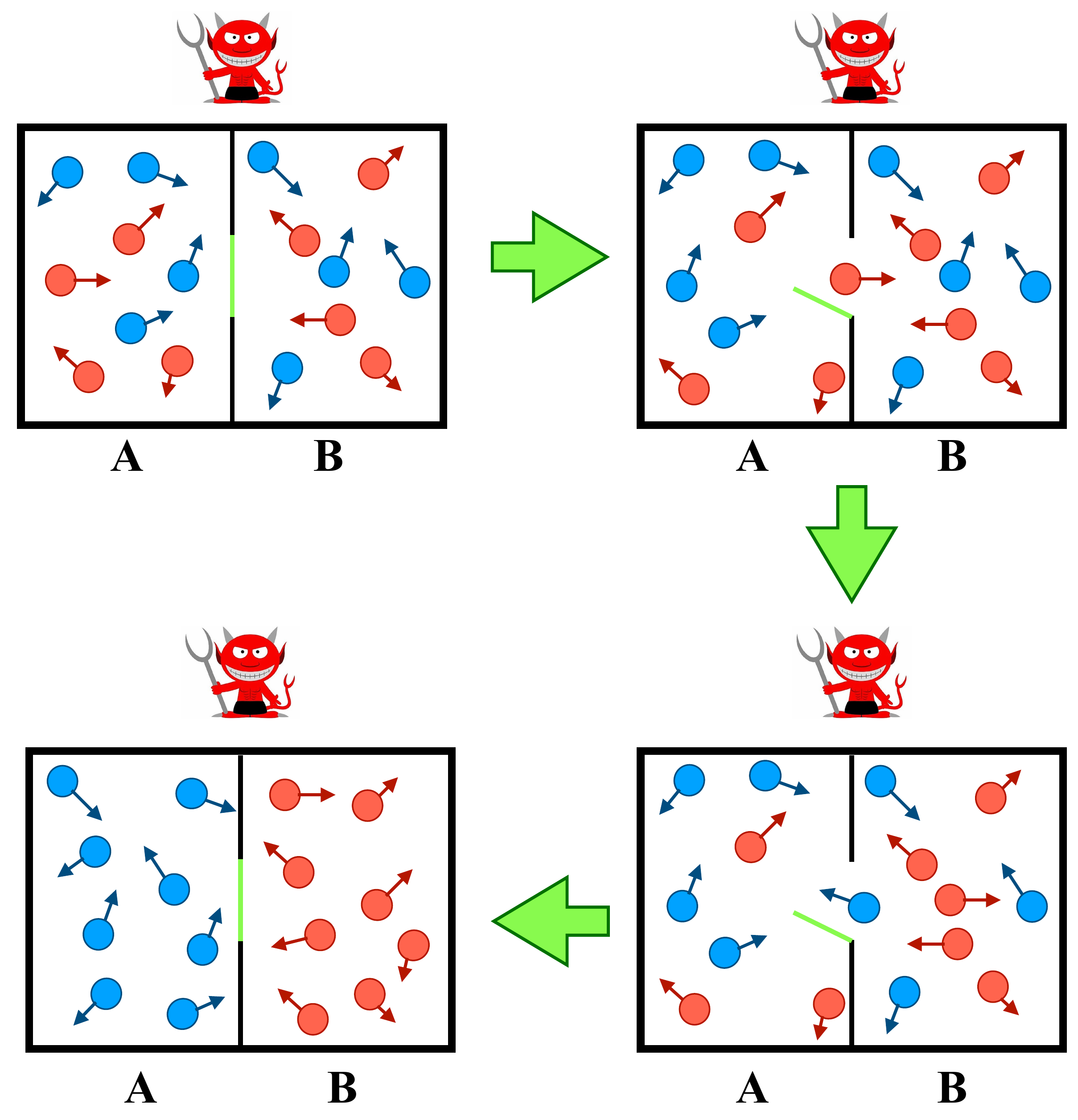}
\caption{ Schematic diagram of Maxwell's demon. The blue coloured circles represent the  molecules with velocities slower  than the average velocity while the red circles represent the faster - than - average molecules.}
\label{maxwells_demon}
\end{figure}     

In 1871, Maxwell proposed a thought experiment that allows this situation - forbidden by the second  law - to happen. He considered a box filled with gas at uniform temperature. Though the mean velocity of any great number of molecules of the gas, arbitrarily selected, 
is almost exactly uniform, the velocities of each individual molecule are by no means uniform. Now, the box is divided into two parts, $A$ and $B$(see Fig.\ref{maxwells_demon}). Maxwell suggested a `demon', i.e., an intelligent being, that can observe the molecules, is placed near the trapdoor 
between the two parts. When a faster - than - average molecule from $A$ flies towards the trapdoor, the demon allows the molecule to pass from $A$ to $B$. On the other hand when a  slower - than - average molecule flies from $B$ towards the trapdoor, the 
demon allows it to pass from $B$ to $A$. The demon doesn't allow slower - than - average molecules to pass from $A$ to $B$ and  faster - than - average molecules to pass from $B$  to $A$. As a result, after some time the average speed of the molecules in 
$B$ would increase while it would decrease in $A$. According to the kinetic theory of gases, average molecular speed is proportional to the temperature. This implies that the temperature decreases in $A$ and increases in $B$, in contrary 
to the second law of thermodynamics. Now one can use $A$ and $B$ as heat baths at different temperatures and can extract work, making it work like a heat engine.       	 
 
 \section{Szilard Engine (SZE)}
\begin{figure}[h]
 \vspace{0.5cm}
 \centering
\includegraphics[width=10 cm]{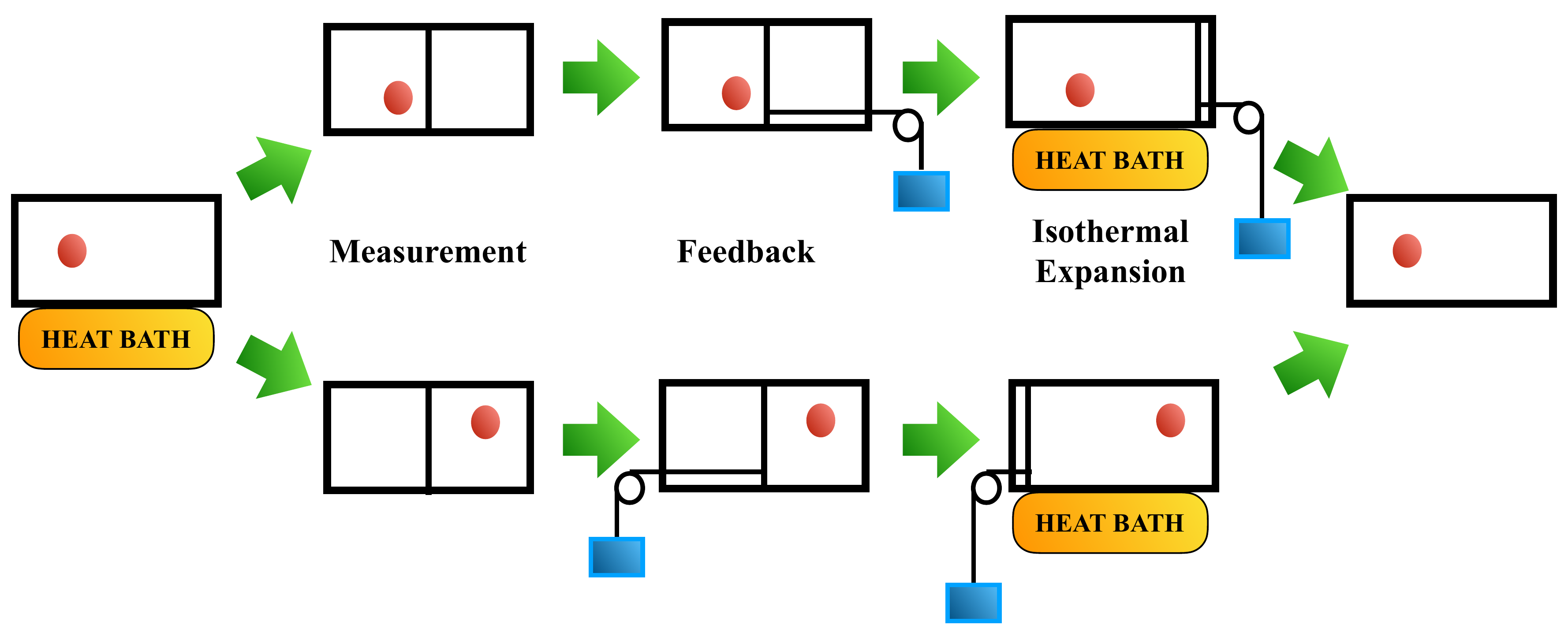}
\caption{ Schematic diagram of Szilard engine.}
\label{szilard_engine}
\end{figure}

A classical demonstration of Maxwell's demon was put forward by Szilard in another thought experiment\cite{szilard}(Fig.\ref{szilard_engine}). \textit{Szilard's engine} contains a single classical particle, obeying the equation of state of an ideal gas, confined in a box of volume $V$. The box is in contact with a thermal reservoir at temperature $T$. The particle thermalises after every collisions with the walls of the box. After some time, a demon inserts a moveable wall in the middle of the box and measures the position of the particle. Depending on the measurement outcome a load is attached to the system. The one particle gas is then allowed to expand reversibly. At the end, the piston is removed and allowed to thermalise, thereby retaining its original state.  

The energetics of the Szilard engine gives insight about the relationship between thermodynamics and information. We can assume that insertion and removal of piston can be done reversibly and hence no energy is expended. The only energetic contribution arises from the reversible expansion of the particle against the piston from $V/2$ to $V$. The work done by the piston on the particle is 
\begin{align}
W=-\int_{V/2}^V pdv=-k_BT\int_{V/2}^V\frac{dV}{V}=-k_BT\ln 2.
\label{w_half}
\end{align}      
In the next subsections we will discuss more generalised situations where we incorporate the following facts: (a) the partition is placed at an arbitrary position inside the box, (b) measurement errors and (c) multi-particle case.

 \subsection{Single particle SZE}
 \label{single_sze}
 A single particle classical gas is confined in a box of length $L$, surface area $A$ and hence volume $V=AL$. The gas is initially at thermal equilibrium with a single thermal bath at inverse
temperature $\beta$. The cyclic process for the SZE is carried out in the following four steps:\\
{\it Step 1. Insertion of partition}: We insert a partition inside the box at position that divides the box in the ratio $x:1-x(0\leq x\leq 1)$. Due to the partition the particle can be on the left 
side or right side. Let's denote the position of the particle by $r$. For simplicity we take $r=1$ when the particle is in left and $r=0$ when it is in the right side. It is to be noted that since the
initial partitioning of the box is unequal i.e., $x\neq 0.5$, the probability of the particle to be in either side is not same. In fact the probability of the particle to be on the left  $P(1)=x$ 
whereas the probability of the particle to be in the right side  $P(0)=1-x$.  We do not need to do any work in this step. \\
{\it Step 2. Measurement}: Next we measure the position of the particle. The measurement outcome, denoted by $q$, can also take two values i.e., 1 or 0. After measurement, $(r,q)$ represents a particular configuration depicting the actual position of the particle and measured outcome. In case of error free measurement $r=q$ and hence we will have only two configurations. Configurations corresponding to erroneous measurement and its effect on work extraction is discussed later. \\
{\it Step 3. Feedback}: We next move the partition quasistatically and isothermally depending on the measurement outcome $q$. If $q=0$ i.e., the particle is on the left side, the partition is shifted
to the right end of the box. If $q=1$, the partition is shifted to the left end. In this process we extract work from the engine. Since the process is quasistatic, the work extraction is given by 
the change in free energy ($\Delta F$) of the particle i.e., $W=\Delta F=(1/\beta)\ln[Z(X_2)/Z(X_1)]$, where $X$ is the position of the partition with $X_1$ and $X_2$ being the initial and the final 
position respectively. $Z(X)$ is the partition function of the single particle system with the partition at $X$ and it is proportional to $X$. Let $W^{(r,q)}$ denote the work for a particular 
configuration. Hence,
\begin{equation}
 \beta W^{(0,0)}=-\ln x;\beta W^{(1,1)}=-\ln (1-x)
\end{equation}
\\
{\it Step 4. Removal of partition}: We remove the partition without any work and the engine returns to its initial state.\\
From the total process, we extract the average work
\begin{align}
\boxed{\beta\la W_{ext}\ra=-x\ln x-(1-x)\ln(1-x)}.
\end{align}

When the partition is placed in the middle, i.e., $x=1/2$ case, using the above expression  we get back Eq.\ref{w_half}.
\subsubsection{Measurement error in single particle SZE}
\label{err_nint_s}
\begin{figure}[h]
 \vspace{0.5cm}
\centering
  \includegraphics[width=7cm]{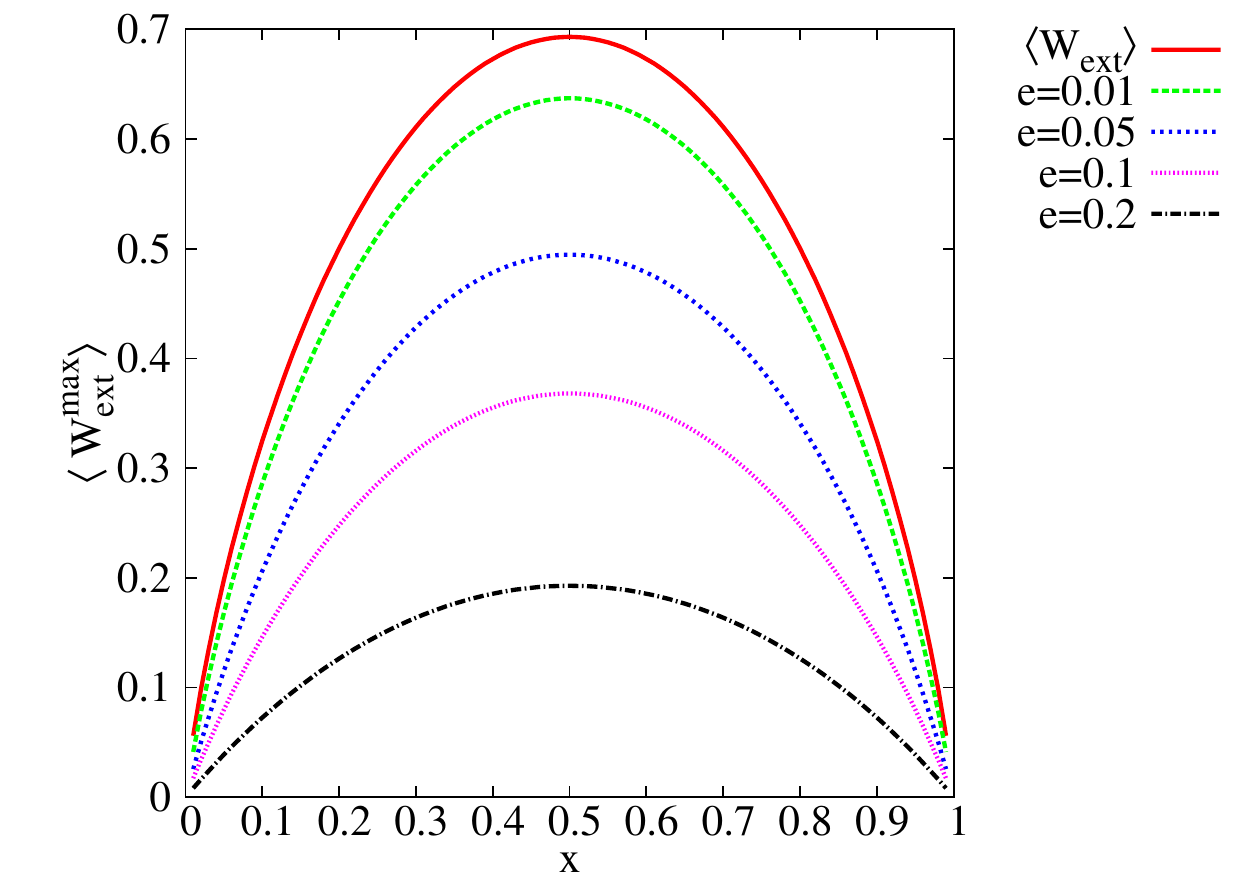}
  \caption{ Plot of maximum average work extracted from single particle SZE as a function of the initial position of the partition for different measurement errors.}
   \label{WI_vs_x}
\end{figure}

Any measurement process can be accompanied by an error. The measurement error is characterized by the 
conditional probabilities $P(q|r)$ - the probability of getting an outcome $q$ given that the actual particle position is $r$. If $e$ is the error rate, then $P(0|0)=P(1|1)=1-e$
and $P(0|1)=P(1|0)=e$. Now $(r,q)$=(0,0), (0,1), (1,0), (1,1) will give four possible configurations of the SZE after measurement.
\begin{figure}[h]
  \vspace{0.5cm}
 \centering
  \includegraphics[width=7 cm,height=10 cm]{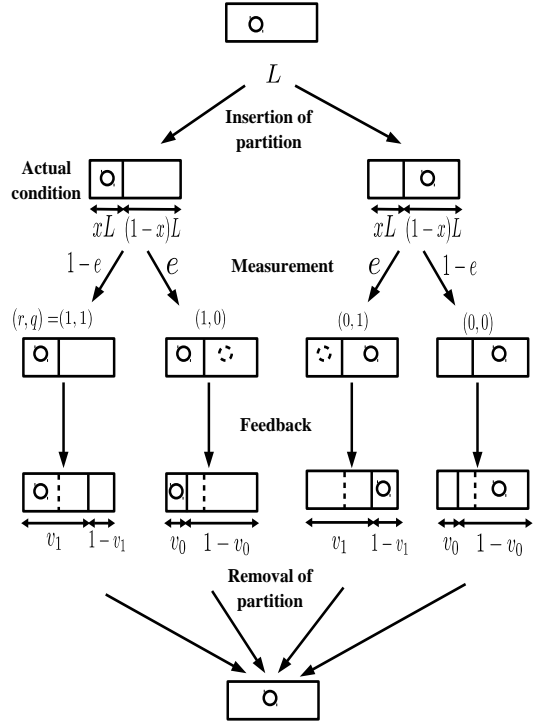}
  \caption{ Single particle SZE. $e=0$ implies error free measurement and in that case only two configurations are possible namely $(r,q)=(1,1)$ and $(0,0)$\cite{sagawa_12}.}
  \label{single}
 \end{figure}
The average information obtained due to measurement is quantified in the following way:
\begin{equation}
 \la I\ra=\sum_{r,q=0,1}\tilde P(r,q)\ln\left[\frac{\tilde P(r,q)}{P(r)P_m(q)}\right],
\end{equation}
where $\tilde P(r,q)=P(q|r)P(r)$ is joint probability of actual and measured position of the particle. $P_m(q)=\sum_{r=0,1}\tilde P(r,q)=\sum_{r=0,1}P(q|r) P(r)$ is the probability of a particular
measured outcome $q$. Using these expressions one can easily calculate that
\begin{eqnarray}
 P_m(1)&=&x(1-e)+(1-x)e,\nn\\
 P_m(0)&=&xe+(1-x)(1-e),\nn
\end{eqnarray}
and hence the average information gain is given by 
\begin{eqnarray}
\la I\ra&=&e\ln e+(1-e)\ln(1-e)\nn\\
&-&[x(1-e)+(1-x)e]\ln[x(1-e)+(1-x)e]\nn\\
&-&[xe+(1-x)(1-e)]\ln[xe+(1-x)(1-e)].\nn\\
\end{eqnarray}
Since the feeback step depends highly on the measurement outcome, it will also be modified due to the presence of error during measurement step. We assume that, after we move the partition the 
volume of the box is  divided in the ratio as $v_1:1-v_1$  for $q=1$ and $v_0:1-v_0$  for $q=0$ with $0\leq v_0,v_1\leq 1$. In this process we extract work from the engine. Let $W^{(r,q)}$ denote the 
work for a particular configuration. Hence
\begin{eqnarray}
&& \beta W^{(1,1)}=\ln\left(\frac{v_1}{x}\right);\beta W^{(1,0)}=\ln\left(\frac{v_0}{x}\right),\nn\\
&& \beta W^{(0,1)}=\ln\left(\frac{1-v_1}{1-x}\right);\beta W^{(0,0)}=\ln\left(\frac{1-v_0}{1-x}\right).\nn
\end{eqnarray}
 The average extracted work in the entire process is given by
\begin{eqnarray}
 \beta\la W_{ext}\ra&=&-x\ln x-(1-x)\ln(1-x)\nn\\
 &&+(1-e)[x\ln v_1+(1-x)\ln(1-v_0)]\nn\\
 && +e[x\ln v_0+(1-x)\ln(1-v_1)].
\end{eqnarray}
We then maximize  $\la W_{ext}\ra$ under a given measurement error $e$ and initial position $x$ of the partition by changing $v_1$ and $v_0$. The maximum value of $\la W_{ext}\ra$ is acheived
when
\begin{eqnarray}
 v_1&=&\frac{x(1-e)}{(1-e)x+(1-x)e},\\
 v_0&=&\frac{(1-x)(1-e)}{(1-x)(1-e)+xe}.
\end{eqnarray}
The maximum work is calculated using the above expressions and is given by the information gained due to measurement 
\begin{align}
\boxed{ \beta\la W_{ext}^{max}\ra=\la I\ra} ,
\end{align}
which is the modified second law in presence of feedback applied to a cyclic process. In this particular case it possible to convert full information into work.
Fig.\ref{WI_vs_x} shows the maximum work that can be extracted from the system in presence of errors during measurement procedure. Larger the error lesser the work that is extracted. Work extraction 
is always degraded due to the presence of error as is seen from the same figure.

\subsection{Multi-particle SZE}
\label{multi_sze}
In multi-particle SZE the thermodynamic cyclic consists of the same four steps: insertion of partition, measurement, feedback and removal of partition. The multi-particle SZE consists of
$N$ number of particles as working substance which is contained in a box of length $L$, surface area $A$ and hence volume $V=AL$. The partition is placed initially at a position that divides the box in the 
ratio $x:1-x$. Let there are $m$ particles on the left side and $N-m$ particles on the right. In the next step the number
of particles on the left side is measured. If the measurement procedure is error free, then the device will exactly measure $m$ particles on the left. After that an isothermal expansion is performed
in contact with a heat bath at inverse temperature  $\beta$, where the partition shifts to a position $l_m$ determined by force balance on both sides of the partition. Finally the wall is removed to 
complete the cycle. The average work done by the SZE is given by \cite{kim_11}
 \begin{align}
 \boxed{\la W_{ext}\ra=-\frac{1}{\beta}\sum_{m=0}^N p_m\ln\left(\frac{p_m}{f_m}\right)}.
  \label{w_tot}
 \end{align}
Here $p_m$ and $f_m$ are given by
 \begin{equation}
 p_m=\frac{Z_{m,N-m}(xL)}{\sum_{m'}Z_{m',N-m'}(xL)},
 \label{pm}
 \end{equation}
\begin{equation}
 f_m=\frac{Z_{m,N-m}(l_m)}{\sum_{m'}Z_{m',N-m'}(l_m)},
 \label{fm}
\end{equation}
where $Z_{m,N-m}(X)=Z_m(X)Z_{N-m}(L-X)$ is a partition function that describes the situation of $m$ particles to the left of the partition and the remaining $N-m$ to  the right, in a 
thermal equilibrium. $Z_m(X)=cX^m/m!$ is the partition function of a system consisting of $m$ particles in a box of length $X$ and $c$ is some constant. Physically, $p_m$ denotes the probability that 
there are $m$ particles to the left after partition and $f_m$ represents the probability to choose the case of $m$ particles on the 
left side of the wall when the wall is inserted at $l_m$ in the time backward process. Although Eq. \ref{w_tot} is derived for QSZE, surprisingly, it also holds true for work done by classical SZE 
derived using classical non-equilibrium thermodynamics. However, for both the cases the partition function differ and so does the amount of work. 

\begin{figure}[h]
   \vspace{0.5cm}
  \centering
  \includegraphics[width=7 cm]{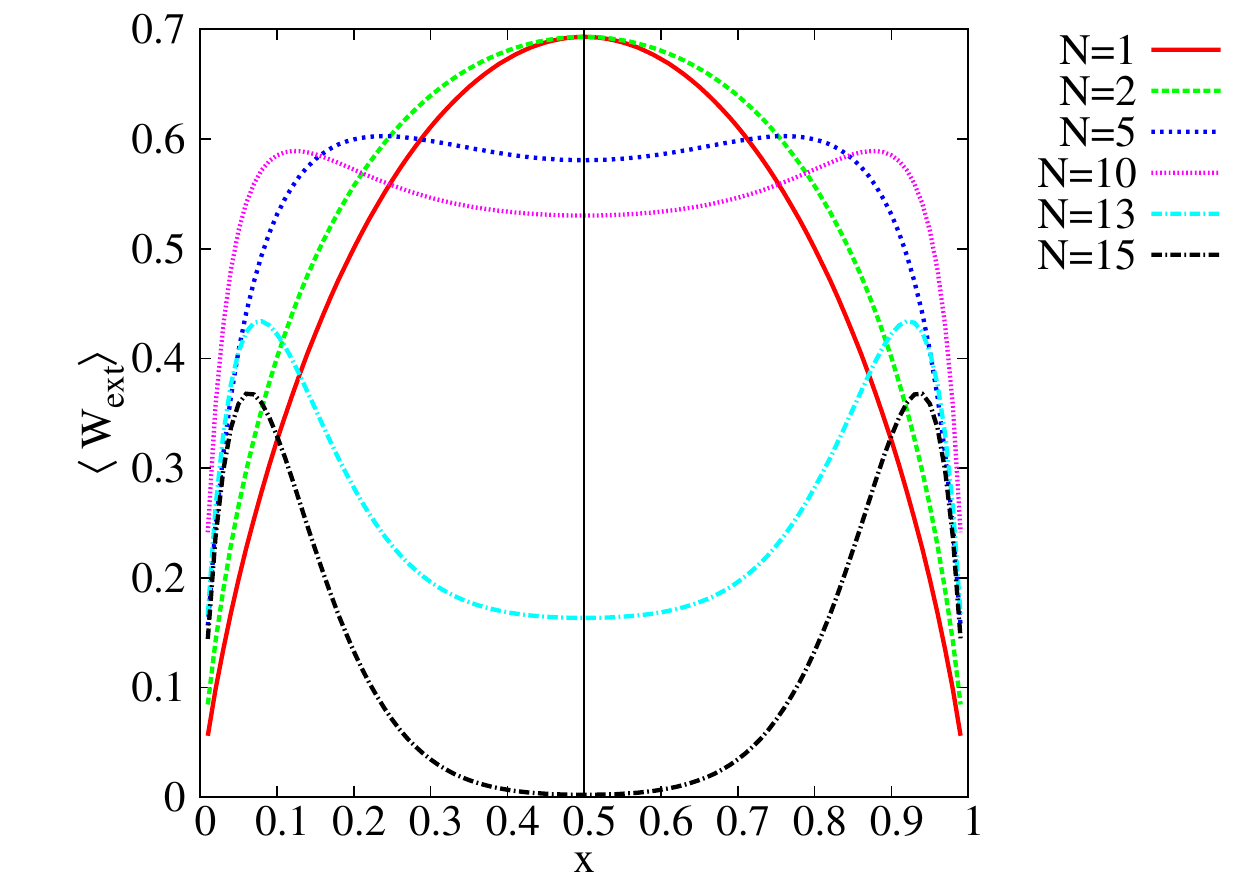}
   \caption{ Plot of average work done by multi-particle SZE as a function of the initial position of the partition for different number of particles.}
   \label{W_vs_x}
 \end{figure}

\begin{figure}[h]
  \vspace{0.5cm}
 \centering
  \includegraphics[width=7 cm]{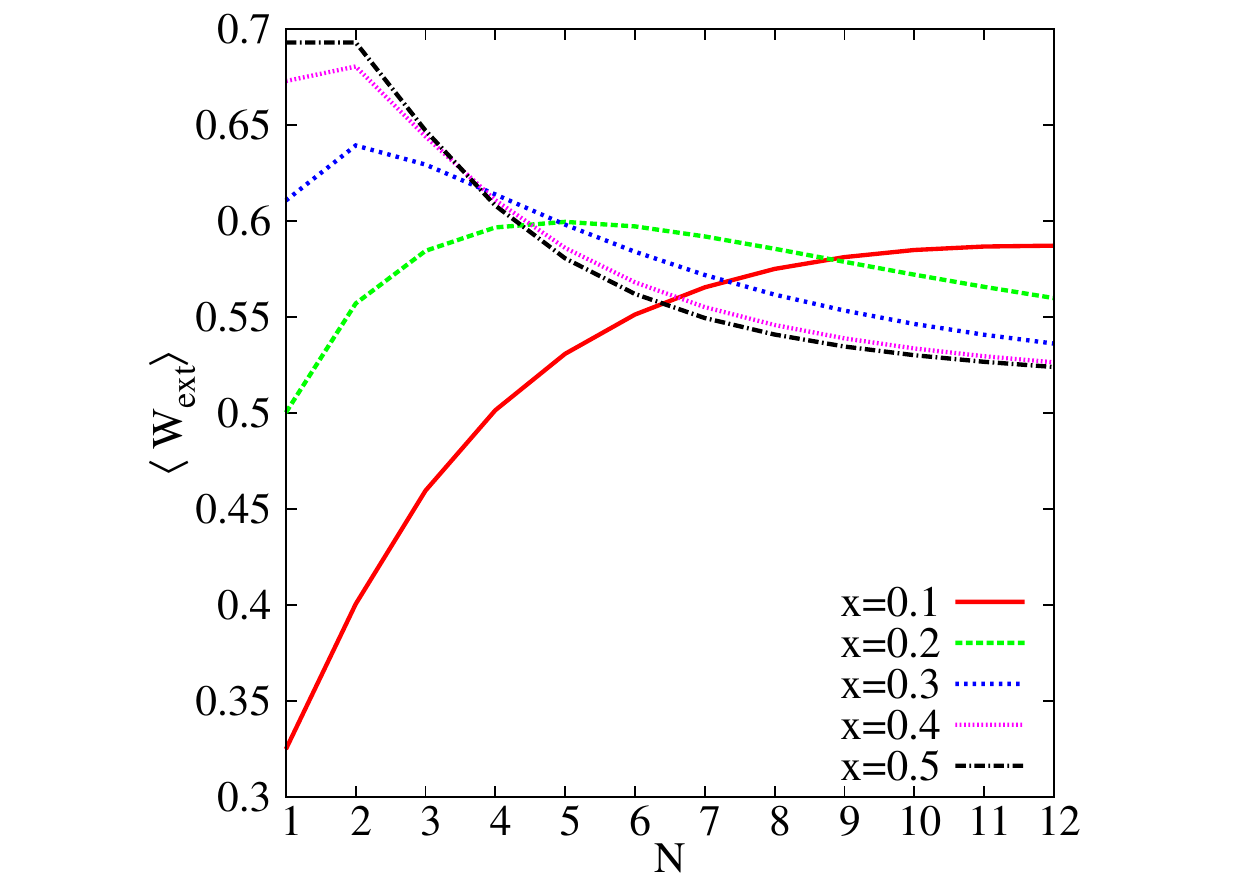}
  \caption{ Plot of average work extraction as a function of number of particles in multi-particle SZE for different initial positions of the partition.}
    \label{W_vs_N}
 \end{figure}

In case of system consisting of $N$ particles, $p_m={N \choose m} x^m(1-x)^{N-m}$ and $l_m=(m/N)L=\alpha L$. $f_m$ is calculated in the following way
\begin{eqnarray}
f_m&=&\frac{Z_{m,N-m}(l_m)}{\sum_{m'}Z_{m',N-m'}(l_m)}, \nn\\
&=&\frac{Z_m(\alpha L)Z_{N-m}((1-\alpha)L)}{\sum_{m'}Z_{m'}(\alpha L)Z_{N-m'}((1-\alpha)L)},\nn\\
&=&\frac{\frac{(\alpha L)^m}{m!}\cdot\frac{[(1-\alpha)L]^{N-m}}{(N-m)!}}{\sum_{m'}\frac{(\alpha L)^{m'}}{m'!}\cdot\frac{[(1-\alpha)L]^{N-m'}}{(N-m')!}},\nn\\
&=&{N \choose m}\frac{(\alpha L)^m[(1-\alpha)L]^{N-m}}{L^N}\nn\\
&=&{N \choose m}\alpha^m(1-\alpha)^{N-m}.
\end{eqnarray}
Inserting  these expressions of $p_m$ and $f_m$ in Eq.\ref{w_tot}, we obtain 
\begin{align}
 \boxed{\la W_{ext}\ra =-T\sum_{m=0}^N {N \choose m} x^m(1-x)^{N-m}\ln\left[\left(\frac{x}{\alpha}\right)^m \left(\frac{1-x}{1-\alpha}\right)^{N-m}\right]}.
 \label{w_tot1}
\end{align}
Fig.\ref{W_vs_x} depicts the behavior of work done by SZE as a function of biasing for different number of particles. It is seen right away that the work extraction is symmetric about $x=0.5$, which  
is quite obvious. Another interesting fact to note is the work done by SZE is same for $N=1$ and $N=2$ when the wall is initially kept at $x=0.5$; it is different otherwise. This is due to the fact that
when there is no biasing, the measurement outcome that there is one particles on both sides does not contribute to the work extraction as the system stays at equilibrium with the wall at $x=0.5$ 
and it does not move. This is not case for $x\neq 0.5$ because in this case even though there is one particle on each side the initial position of the partition is not the equilibrium position.
Work extraction has a single peak at $x=0.5$ for single particle but it splits into two distinct peaks as the number of particles increases. 
Due to the emergence of two peaks a local minima in work extraction is visible at $x=0.5$. This minimum value eventually goes to zero as is evident from Fig.\ref{W_vs_x}.  
This gives clear clue that work extraction is more for multiparticle working substance with biasing.
Fig.\ref{W_vs_N} shows the work extraction as a function of number of particles for different biasing. The plot depicts the fact that work extraction is small with large biasing when number of
particles is small. But work done by SZE increases for large biasing when the particle number is large.

\subsubsection{Measurement error in Multi-particle SZE}
\label{err_nint}
In multi-particle SZE, the error enters in the measurement step where the measuring device can make an error while counting the number of particles on each sides.
\begin{figure}[]
 \vspace{0.5cm}
 \centering
  \includegraphics[height=9 cm,width=7 cm]{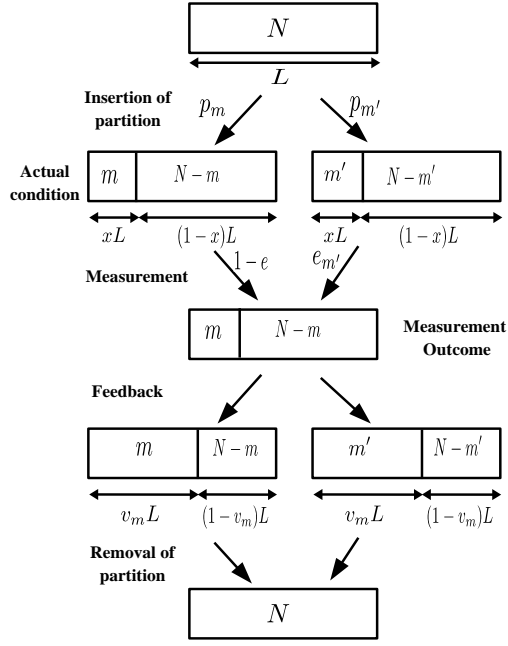}
  \caption{ Multi-particle SZE with errors.}
    \label{multi_particle}
 \end{figure}
In case of $N$ identical particle system there can be $N+1$ configurations. Each configuration gives the number of particles on the left and the right side of the box.
Consider a measurement outcome with the configuration: $m$ particles on the left and $N-m$ particles on the right as shown in Fig.\ref{multi_particle}. The actual configurations can be (i) $m$ 
particles on the left side and $N-m$ particles on the right and (ii) $m'(\neq m)$ particles on the left and $N-m'$ particles on the right. The  probability of observing measured configuration from
the former case is $1-e$. Remaining $N$ actual configurations can be mapped to the measured configuration due to measurement error. This error will depend on the number of particles on the 
left in the actual configuration. Hence the  probability of observing measured configuration from the latter case is $e_{m'}$. Due to normalization of probability
$\sum_{m'\neq m}e_{m'}=e$.  These possibilities are clearly shown in Fig.\ref{multi_particle}. Depending on the outcome, 
the partition is shifted to  a new position where it divides the box in the ratio $v_m:1-v_m$. If $W^{(m,m)}$ represents the work done when the actual and the measured number of particles on the left 
side are same (say $m$) and $W^{(m',m)}$ represents the work done when the actual number of particles is different from the measurement value (say $m'$ and $m$ respectively), then
\begin{eqnarray}
 \beta W^{(m,m)}&=&\ln\left[\frac{Z_{m,N-m}(v_mL)}{Z_{m,N-m}(xL)}\right],\nn\\
 \beta W^{(m',m)}&=&\ln\left[\frac{Z_{m',N-m'}(v_mL)}{Z_{m',N-m'}(xL)}\right].
 \label{w_mm}
\end{eqnarray}
The average work output given a particular measurement outcome $m$ is given by
\begin{align}
\boxed{ \beta \la W^{(m)}\ra=p_m(1-e)\beta W^{(m,m)}+\sum_{m'\neq m}e_{m'} p_{m'}  \beta W^{(m',m)}}.
 \label{w_m}
\end{align}
One can assume the error $e_{m'}$ to be same for all the $N$ configurations. In that case $e_{m'}=e/N$ should be used in Eq.\ref{w_m}.
The total average work extraction is given by
\begin{align}
 \boxed{\beta \la W_{ext}\ra=\sum_{m=0}^N \beta \la W^{(m)}\ra}.
 \label{w_tot2}
\end{align}
Work extraction depends crucially on the nature of the working substance. In case of ideal gas there is no inter-particle interaction and the partition function of $n$-particle system confined in a 
box of dimension $X$ is $Z_n(X)=cX^n/(n!)$. Using this expression of partition function, Eq.\ref{w_mm} can be easily calculated 
\begin{eqnarray}
  \beta W^{(m,m)}&=&\ln\left[\left(\frac{v_m}{x}\right)^m\left(\frac{1-v_m}{1-x}\right)^{N-m}\right],\nn\\
\beta W^{(m',m)}&=&\ln\left[\left(\frac{v_m}{x}\right)^{m'}\left(\frac{1-v_m}{1-x}\right)^{N-m'}\right].\nn\\
  \end{eqnarray}
The above two expressions can be used to calculate the total average extracted work given by Eq.\ref{w_tot2}.
\section{Landauer principle}
The thermodynamic process of Szilard engine is cyclic because the initial and the final states of the particle  are identical and hence the free energy change of the system is zero. This implies that the particle is systematically extracting work cyclically from the thermal reservoir in contrary to the second law of thermodynamics. The resolution of this apparent violation of second law was given by Charles Bennett\cite{bennett}. He investigated the energetics of Szilard's engine using Landauer principle that relates \textit{logical irreversibility} with energy consumption during computing process. The information obtained from measurement is stored in physical systems. As a result simple logical operations is accompanied by an energetic cost. In case of Szilard's engine, the particle in the box might have gone through a cyclic process but the whole system consisting of the particle in the box and a memory (which is used to store  the information of the position of the particle after the piston is inserted) did not complete a whole cycle. The full cycle will be completed  if at the  end when the  piston is removed,  the data in the memory is erased (or reset). This erasure of a bit is a logical irreversible operation, since a bit that is initially in one of the two possible states (0 or 1) is set to a reference state irrespective of its initial state. \textit{Landauer's principle}\cite{landauer_61,landauer_91} states that erasure of a bit requires at least $k_BT\ln2$ amount of work dissipated to the thermal environment
\begin{align}
W_{\text{eras}}\ge k_BT\ln2,
\end{align}
where the equality is satisfied in case of quasistatic erasing process. Landauer's principle was recently tested experimentally in \cite{berut12} using optically-trapped Brownian particles. It is to be noted in this context that error-free measurement process is a \textit{logically reversible} process because there is a one-to-one correspondence between the state of the system and the measured outcome. 

Taking Landauer's principle into account, it is now very clear that in every cycle there are two contributions to the total work: (i) the work done in the system due to the reversible expansion of the one molecule ideal gas and (ii) work due to erasure of measured data. Since the measurement has only two outcomes, the measurement of Szilard's engine is equivalent to measurement of the state of one bit. In every cycle, the state of the demon has to be reset to the reference state in order to carry out the measurement procedure in the next cycle. This resetting process requires a minimum of $k_BT\ln2$ amount of work as stated by Landauer. Hence, the work per cycle of the Szilard engine is
\begin{align}
W_{\text{tot}}=W+W_{\text{eras}}\ge- k_BT\ln2+ k_BT\ln2=0,
\end{align}
which is in accordance with the second law of thermodynamics. Before Landauer's principle, Brillouin, on the basis of a specific model, argued that measurement process requires work which is compensated by the work extracted by the Maxwell's demon\cite{brillouin}. Later Bennett proposed a model in which measurement process can be performed without any work. On the grounds of Landauer's principle and using a specific model, Bennett showed that erasing of memory of the demon is the resolution 
to Maxwell's demon\cite{bennett}.

\begin{figure}[h]
 \vspace{0.5cm}
 \centering
  \includegraphics[height=2 cm,width=10 cm]{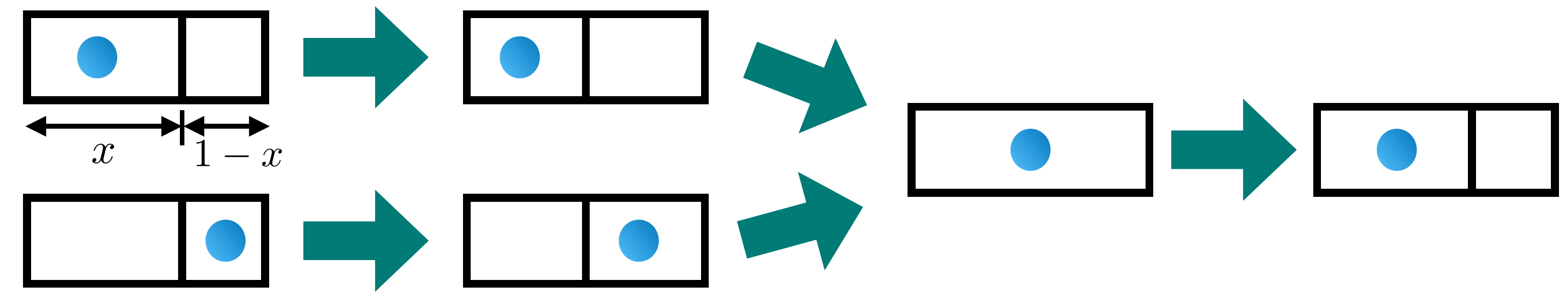}
  \caption{ Schematic picture of information erasure from an asymmetric memory\cite{sagawa_09}.}
    \label{erasure}
 \end{figure}

In the recent years, Landauer's principle, which is based on thermodynamic energy cost of information erasure, has been challenged. A generalised version of Landauer's principle has been put forward by Sagawa and Ueda \cite{sagawa_09}, that sets a lower bound on the energy cost of measurement and information erasure combined. The analysis by Sagawa and Ueda crucially depends on how one construct the memory where the information is stored. Instead of giving a rigorous proof, following \cite{sagawa_09} here we illustrate this idea using an example of a particle in a box with a partition as a model for a memory. We construct the memory in such a way that the partition divides the box in the volume ratio of $x:1-x$. When the particle is in the ``left" (or ``right") side of the box, the memory registers ``0" (or ``1").  Let's consider the initial probability of the outcomes of ``0" or ``1" is equal i.e., $1/2$. Now we will consider the case of quasistatic information erasure from the memory in contact with a heat bath at inverse temperature $\beta$ and calculate the work required for such a process. Erasure of memory is done in three sub-processes. At first the information is stored in the memory (Fig. \ref{erasure}). 
Then partition is then moved to the centre of the box. The average work cost for this is process is given by the free energy change i.e., 
\begin{align}
\beta W^{(1)}_{\text{eras}}&=\frac{1}{2}\left[\ln x-\ln(1/2) \right]+\frac{1}{2}\left[\ln (1-x)-\ln(1/2) \right]\nn\\
&=\frac{1}{2}\left[\ln(2x)+\ln \{2(1-x)\}\right]
\end{align}  
The partition is then removed and this process can be regarded as the free expansion of the gas that requires no work. Finally the box is compressed so that the memory returns to its standard state of ``0". This sub-process occurs at the cost of work given by
\begin{align}
\beta W^{(2)}_{\text{eras}}=-\ln x.
\end{align}
The total work required for information erasure is 
\begin{align}
\beta W^{\text{M}}_{\text{eras}}=\beta W^{(1)}_{\text{eras}}+\beta W^{(2)}_{\text{eras}}=\ln 2-\frac{1}{2}\ln\left(\frac{x}{1-x}\right).
\end{align}
In case of symmetric memory we have $x=1/2$ and the work required for information erasure reduces to $\ln 2$ - which is the limit set by Landauer erasure principle. For any other arbitrary values of $x$, Landauer bound is not followed.
\begin{figure}[h]
 \vspace{0.5cm}
 \centering
  \includegraphics[height=1.5 cm,width=9 cm]{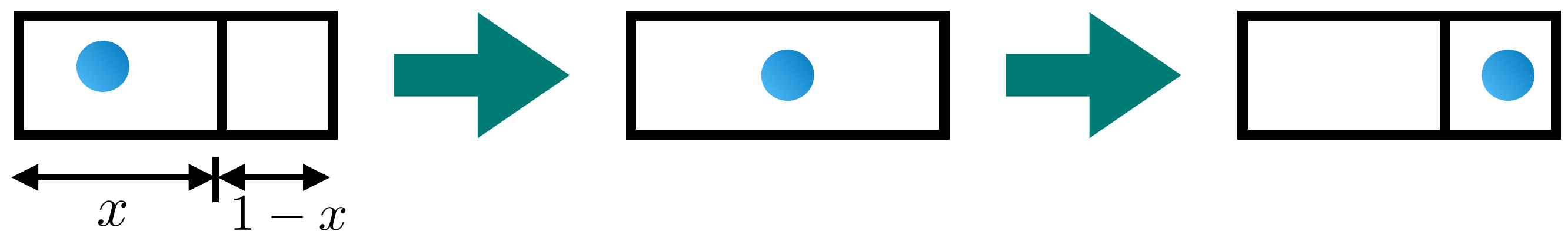}
  \caption{ Schematic picture of the process of information writing  to an asymmetric memory\cite{sagawa_09}.}
    \label{measurement}
 \end{figure}
We now consider a quasistatic measurement process in presence of a heat bath at temperature $T$(Fig.\ref{measurement}). Let the memory is initially in the standard state ``0". If the measurement outcome is ``0", the state of the memory does not change. If the measurement outcome is ``1", then the left side of the box expands and pushes the box to the right end which requires $(1/\beta)\ln x$ amount of work. The box then compresses from the left side until the partition comes back to its initial position. This compression process requires $-(1/\beta)\ln (1-x)$ amount of work. Since the probability of measurement outcome ``0" or ``1" is same i.e., $1/2$,  the average energy cost for the measurement procedure is given by
\begin{align}
\beta W^{\text{M}}_{\text{meas}}=\frac{1}{2}\ln\left[\frac{x}{1-x}\right].
\end{align}
One can make the above work zero by considering a symmetric($x=0.5$) memory in classical systems. 
The total work required for information measurement and erasure process combined is
\begin{align}
\beta W^{\text{M}}=\beta W^{\text{M}}_{\text{meas}}+\beta W^{\text{M}}_{\text{eras}}=\ln 2.
\end{align}
The above expression by Sagawa and Ueda\cite{sagawa_09} unifies the approaches adopted by Brillouin, Landauer and Bennett, and showed that the total work due to measurement and erasure process compensates for the work extracted by the Maxwell's demon.
\section{Conclusion}
In this article, we have given a flavour of the initial building stages of  information thermodynamics by discussing three important works in this field namely Maxwell's demon, Szilard engine and Landauer principle. A nice future prospect in this area is to investigate these works in interacting multi-particle systems.  It is important to note that all the works has been discussed using the concepts of classical mechanics. In recent years these works has been extended to quantum regime and a lot of work is being carried out in quantum thermodynamics. Specifically, Kim et al.\cite{kim_11} has given a theoretical analysis of quantum version of Szilard engine where they studied have studied Szilard engines consisting of Bosons and Fermions. They pointed out that, unlike classical SZE, the insertion and removal of partition in the quantum version of Szilard engine requires work due to change change in the energy levels on the either side of the partition.  More (less) work (with respect to classical Szilard engine) can be extracted from the Bosonic(Fermionic) Szilard engine due to the indistinguishability of identical particles.

\section{Acknowledgment}
Arun M Jayannavar thanks DST, India for financial support (through J. C. Bose National Fellowship).


\end{document}